\documentclass[conference,9pt]{IEEEtran}
\IEEEoverridecommandlockouts

\usepackage{cite}
\usepackage{amsmath}
\usepackage{amssymb}
\usepackage{graphicx}
\graphicspath{{images/}}
\usepackage[dvipsnames,table]{xcolor}
\usepackage{microtype}

\usepackage{array}
\usepackage{arydshln}
\usepackage{multirow}
\usepackage{booktabs}
\usepackage{threeparttable}
\usepackage{siunitx}

\usepackage[caption=false,font=footnotesize]{subfig}

\usepackage{algorithm}
\usepackage{algpseudocode}

\usepackage{xspace}
\usepackage[normalem]{ulem}
\usepackage{pifont}

\usepackage[varqu,scaled=0.95]{zi4}
\usepackage{listings}

\usepackage[hidelinks]{hyperref}

\algdef{SE}[PARFOR]{ParFor}{EndParFor}[1]{\textbf{parallel for}\ #1\ \textbf{do}}{\textbf{end parallel for}}
\algnewcommand{\Let}[1]{\State \textbf{let}\ #1}
\algrenewcommand\algorithmiccomment[1]{\hfill\textit{\small // #1}}

\definecolor{progaie}{HTML}{1F5FBF}
\definecolor{progauto}{HTML}{C55A11}
\definecolor{progq}{HTML}{1F5FBF}
\definecolor{proge}{HTML}{7030A0}
\newcommand{\pc}[1]{\makebox[1.425em][r]{\textcolor{black!50}{#1}}}
\lstdefinelanguage{stageprog}{
  morekeywords={SETB,ASSEMBLE,GEMM,MATTN,CTXUPD,HALT},
  morekeywords=[2]{AIE},
  morekeywords=[3]{AUTO},
  morekeywords=[4]{PL},
  morekeywords=[5]{rms},
  morekeywords=[6]{tri,res,silu,glu,qkv8,argmax},
  morecomment=[l]{;},
  sensitive=true}
\lstdefinestyle{aiecode}{
  language=C++, basicstyle=\ttfamily\scriptsize, keywordstyle=\bfseries,
  commentstyle=\color{black!55}, numbers=none, xleftmargin=0pt, framexleftmargin=0pt,
  frame=tb, framerule=0.4pt, rulecolor=\color{black},
  aboveskip=2pt, belowskip=2pt, framesep=2pt,
  columns=fixed, basewidth=0.475em, keepspaces=true, showstringspaces=false}
\lstdefinestyle{stageprog}{
  language=stageprog,
  basicstyle=\ttfamily\scriptsize,
  keywordstyle=\bfseries,
  keywordstyle=[2]\bfseries\color{progaie},
  keywordstyle=[3]\bfseries\color{progauto},
  keywordstyle=[4]\bfseries\color{black!60},
  keywordstyle=[5]\color{progq},
  keywordstyle=[6]\color{proge},
  commentstyle=\color{black!55},
  numbers=none, xleftmargin=0pt, framexleftmargin=0pt,
  frame=tb, framerule=0.4pt, rulecolor=\color{black},
  aboveskip=2pt, belowskip=2pt, framesep=2pt,
  columns=fixed, basewidth=0.475em, keepspaces=true, showstringspaces=false,
  mathescape=false, escapeinside={(*}{*)}}

\newcommand{\Design}{\textsc{HeteroVLM}\xspace}

\newcommand{\circnum}[1]{\raisebox{.5pt}{\textcircled{\raisebox{-.9pt}{\scriptsize #1}}}}
\newcommand{\reqnum}[1]{\raisebox{-.14em}{\scalebox{1.28}{\ding{\numexpr201+#1\relax}}}}

\hypersetup{
  pdftitle={HeteroVLM: Edge Multi-Request Vision-Language Model Serving on a Heterogeneous Versal SoC with AI Engines and Programmable Logic},
  pdfauthor={Zhiheng Chen, Yu-An Chou, Yifan Zhang, Ye Qiao, Saptarshi Mitra, Sitao Huang},
  pdfsubject={},
  pdfkeywords={},
  pdfcreator={},
  pdfproducer={}
}

\begin{document}
\bstctlcite{IEEEexample:BSTcontrol}

\title{\fontsize{23}{29}\selectfont \Design: Edge Multi-Request Vision-Language\\ Model Serving on a Heterogeneous Versal SoC\\ with AI Engines and Programmable Logic}

\author{\IEEEauthorblockN{Zhiheng Chen\IEEEauthorrefmark{1}, Yu-An Chou\IEEEauthorrefmark{1}, Yifan Zhang, Ye Qiao, Saptarshi Mitra, Sitao Huang}
\IEEEauthorblockA{University of California, Irvine\\
\{zhihenc5, chouy10, yifanz58, yeq6, saptarm, sitaoh\}@uci.edu}
\thanks{\IEEEauthorrefmark{1}These authors contributed equally to this work.}}

\maketitle

\begin{abstract}
Vision-language models (VLMs) play a pivotal role in enabling autonomous vehicles and robots to perceive and reason about the physical world. Edge computers in autonomous systems typically serve concurrent VLM inference requests with various deadlines, each mixing a compute-bound encoder and prefill stage with a memory-bound decode stage. GPU serving optimizes for high throughput in datacenters, with per-iteration host scheduling, which can cause an urgent request to wait for a long time, and small-batch decode costs much energy per token. Prior accelerators on Versal AI Engine (AIE) SoCs map one model onto a fixed AIE dataflow graph for one input stream, without decode or request scheduling, which limits their capability in serving multiple requests. We present \Design, which is optimized for  multi-request VLM serving on a resource-constrained heterogeneous platform. It keeps the AIE graph static and makes it schedulable. The AIE array runs the encoder, prefill and attention, and the programmable logic (PL) implements GEMV engines for small decode steps and a scheduler that orders short work units by class and deadline. Concurrent requests share one decode step that reads the weights once. An urgent request preempts at a work-unit boundary, and the paused request resumes from its results in DDR memory. On an AMD Versal VC1902, text and camera questions get their first tokens in 140 and 303\,ms. \Design decodes a single request 1.3$\times$ and 4.0$\times$ faster than NVIDIA Tesla T4 and Jetson Orin Nano Super GPUs, consuming only 61\% and 73\% of their energy per token, or 51\% of the energy per token of an NVIDIA GeForce RTX 3090 Ti.  A single bitstream runs multiple VLMs, switching between them at run time. A demo is available at \url{https://anonymous.4open.science/r/demo_heterovlm-EFM2/}.
\end{abstract}

\section{Introduction}
\label{sec:intro}

\begin{figure*}[t]
  \centering
  \includegraphics[width=\textwidth]{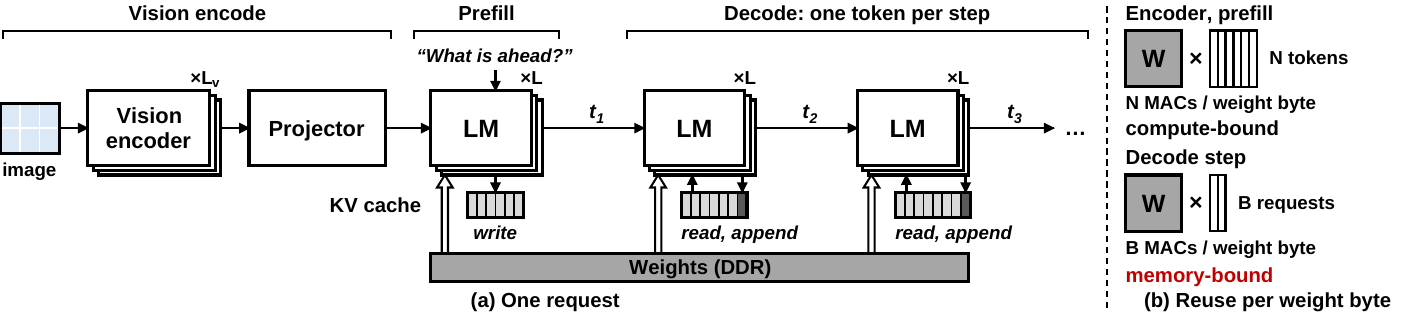}
  \caption{Inference of one VLM request. (a) Encoder and prefill process all tokens in one pass; each decode step re-reads the weights and produces one token. (b) Reuse per weight byte: $N$ tokens in prefill, $B$ requests in decode.}
  \label{fig:vlm}
\end{figure*}

Vision-language models (VLMs)~\cite{lfm2,lfm25vl3b} answer questions about images. In a vehicle or a robot, one edge server runs such a model for several functions at once: a hazard question about the front cameras, a check of the driver, a caption of the surround view and a text question to the assistant can be active together~\cite{drivevlm,cui2024survey,canas2025dms}. Safety questions have deadlines while chat answers can wait, so the server handles concurrent requests with different deadlines. Each request also mixes three workloads (Fig.~\ref{fig:vlm}). An image encoder turns each image into tokens, and the language model reads them with the prompt in one prefill pass and then generates the answer one token per decode step. The encoder and the prefill process hundreds of tokens at once, a few hundred multiply-accumulates (MACs) per byte of DDR traffic, so they need compute. Each decode step re-reads all weights for one token, about one MAC per byte, so decode needs memory bandwidth.

GPU serving systems target datacenter throughput and schedule requests on the host. Decode iterations are often replayed as CUDA graphs to hide launch cost~\cite{pygraph}, so the host acts between iterations, and run-to-completion scheduling makes later requests wait behind running ones~\cite{fastserve}. At edge batch sizes, GPU decode is memory-bound and costs much energy per token.

AMD Versal AIE SoCs pair an array of AI Engines (AIE), dense vector processors that suit the compute-bound phases, with programmable logic (PL). The PL is flexible, with custom datapaths and control for each operation (any data width, fused per-row operations, fine-grained scheduling). Its memory engines at the DDR ports keep the channels busy streaming weights, so FPGA decoders reach most of their DDR bandwidth~\cite{llamakv260,hummingbird}. Prior work on these SoCs builds and composes matrix-multiply engines~\cite{charm,automm}, compiles kernels to them~\cite{aries}, or accelerates transformer encoders such as ViT~\cite{ssr,cat,eqvit}. CHARM~\cite{charm} leaves softmax, LayerNorm and transposes to separate PL kernels that run through DDR between the matrix multiplies. \Design fuses these operators into the PL stages that feed and drain the AIE. These designs map the model onto a dataflow graph fixed at compile time and serve one input stream, as do FPGA language-model accelerators~\cite{llamakv260,tellme,cobra}. None of them runs an autoregressive VLM for concurrent requests with deadlines.

\Design keeps the AIE graph static as a predictable compute engine and makes the graph schedulable. A scheduler in the PL splits each request into short work units, orders them by class and deadline, and lets an urgent request preempt at a work-unit boundary. The paused request then resumes from its results in DDR. The encoder, prefill and attention run on the AIE. Decode steps of fewer than $b_\mathrm{thr}$ requests, a threshold set by the host, run on GEMV engines in the PL, and larger ones run on the AIE GEMM groups once no request waits for its first token. Concurrent requests share one decode step that reads the weights once. Our contributions are:
\begin{itemize}
  \item A complete VLM accelerator design on a Versal AIE SoC (Section~\ref{sec:hardware}). Two AIE GEMM groups and an attention block run the compute-bound work, and the PL holds four decode GEMV engines, the scheduler, the memory servers, quantization, image preprocessing and fused epilogues. The epilogues apply the operations that follow each matrix multiply to its sums as they stream out of the AIE. The same bitstream and AIE graph serve other VLMs of the same shape class, such as LFM2.5-VL-1.6B and SmolVLM2-2.2B, with their own programs and weights (Section~\ref{sec:hw:overview}).
  \item Multi-request execution (Section~\ref{sec:method}) scheduling and optimization. Concurrent requests share decode steps on the PL or the AIE, and an urgent request preempts after the running GEMM work unit and the attention calls already issued.
  \item Complete evaluation on the VC1902 SoC running LFM2.5-VL-3B~\cite{lfm25vl3b} and two other VLMs with concurrent requests, preemption and resumption (Section~\ref{sec:experiments}). It covers latency, throughput, power, design choices and comparisons with prior work and NVIDIA Tesla T4, Jetson Orin Nano Super and GeForce RTX 3090 Ti GPUs. A 60-token text question gets its first token in 140\,ms and a 177-token camera question with one 512$\times$288 image in 303\,ms. Two requests share one PL decode step, doubling throughput. A single request decodes 1.3 and 4.0 times faster than on the T4 and Orin Nano Super, with 61\%, 73\% and 51\% of the energy per token of the T4, Orin Nano Super and RTX 3090 Ti.
\end{itemize}

\section{Background}
\label{sec:background}

\subsection{VLM Inference}
\label{sec:bg_vlm}
Fig.~\ref{fig:vlm}(a) shows one request in an edge-class VLM: a vision encoder, a projector and a language model (LM). In LFM2.5-VL-3B~\cite{lfm2,lfm25vl3b}, for example, a SigLIP2 encoder~\cite{siglip2} of 412M parameters in 27 layers of width 1152 takes $16\times16$-pixel patches, and a projector of 14M parameters merges each $2\times2$ group of patches into one LM token, so 576 patches give 144 tokens. The LM has 2.7B parameters in 30 layers of hidden width 2048: 22 gated short-convolution and 8 grouped-query attention layers.

The encoder and the \emph{prefill} of the LM process the image and all prompt tokens in one pass. The prefill writes the keys and values of the attention layers into a KV cache and yields token $t_1$. Each \emph{decode} step reads the whole cache, appends one entry and yields one token, so the cache grows by one entry per step. The prefill and each decode step read the LM weights again, 2.7\,GB per step with int8 weights. Each weight byte read from DDR thus serves the $N$ tokens of a pass in the encoder and the prefill, which are compute-bound, but only the $B$ requests of a step in decode, which is memory-bound (Fig.~\ref{fig:vlm}(b)).

\subsection{The Versal VC1902 AIE SoC}
\label{sec:bg_platform}
The Versal VC1902 AIE SoC contains 400 first-generation AIE tiles running at 1.25\,GHz, each with a 32\,KB local memory, 64 int16$\times$int8 MACs per cycle and cascade links to its neighbors, with a PL of 900k LUTs and 1,968 DSP58 engines~\cite{amd_ds950,amd_am009,amd_ds957}. On our card, its DDR4 memory delivers about 50\,GB/s memory bandwidth. The array exchanges streams with the PL and runs a static dataflow graph when the device is configured.  Scheduling happens in the PL or on the host.

\subsection{Related Work}
\label{sec:related_work}
\begin{table}[t]
\centering
\caption{Comparison with related accelerators.}
\label{tab:related}
\scriptsize
\setlength{\tabcolsep}{2.5pt}
\renewcommand{\arraystretch}{0.85}\setlength{\aboverulesep}{0.25ex}\setlength{\belowrulesep}{0.4ex}
\begin{tabular}{@{}lcccc@{}}
\toprule
 & \Design & CHARM & CAT & SSR \\
 &  & \cite{charm} & \cite{cat} & \cite{ssr} \\
\midrule
Device & VC1902 & VC1902 & VC1902 & VC1902 \\
Model & LFM2.5-VL, SmolVLM2 & BERT, ViT & BERT, ViT & ViT \\
Parameters & 1.6--3.1\,B & -- & 86--110\,M & 4--7.4\,M \\
Numerics & int8/int14 & fp32 & int8 & int8 \\
Weights in DDR & 1.8--3.4\,GB & streamed & 86--110\,MB & on chip \\
Phases & encoder, prefill, decode & MM on AIE & encoder & encoder \\
Requests & concurrent & batch & batch & batch \\
Preemption & priority, deadline & -- & -- & -- \\
\bottomrule
\end{tabular}
\end{table}
On Versal AIE SoCs, CHARM~\cite{charm} accelerates matrix multiplication on the AIE. Since one large accelerator is poorly used by small layers, CHARM composes several matrix-multiply accelerators of different sizes. Its release runs only the matrix multiplies, and its paper adds PL kernels for softmax, LayerNorm and transposes through DDR. SSR~\cite{ssr} forwards data between such accelerators on chip to trade latency against throughput, and AutoMM~\cite{automm} targets several data types. CAT~\cite{cat}, EQ-ViT~\cite{eqvit} and HG-PIPE~\cite{hgpipe} accelerate transformer encoders such as ViT on the AIE or the PL. Like CHARM, \Design runs matrix multiplies on separate groups of AIE tiles; it fuses the other operators into the PL and adds an attention block, the LM with its decode phase and a scheduler for concurrent requests. Table~\ref{tab:related} compares the features of these designs.

PREMA~\cite{prema} and Planaria~\cite{planaria} bring multi-tenancy and preemption to DNN accelerators for CNN and RNN inference. PREMA preempts a TPU-like NPU in time under a predictive scheduler, and Planaria splits a systolic array in space among co-located DNNs by priority and latency target. \Design applies priority and deadline preemption at work-unit boundaries to a VLM server on a Versal AIE SoC.

Decoders on embedded FPGAs~\cite{llamakv260,hummingbird} reach most of their DDR bandwidth for one request with 4-bit weights. TeLLMe~\cite{tellme} runs LLM prefill and decode with ternary weights, and COBRA~\cite{cobra} runs binary transformers. \Design keeps int8 weights and raises the work per weight byte by letting requests share a decode step.

GPU serving systems target the datacenter, with many GPUs and large batches. Orca~\cite{orca} batches requests per decode iteration, DistServe~\cite{distserve} places prefill and decode on separate GPUs, and EPD disaggregation~\cite{epd} also separates the encoder, all scheduled by host software. \Design applies shared decode steps and phase separation on one chip, scheduled in the PL.

\section{Design Method and Architecture}
\label{sec:hardware}

\subsection{Design Flow and Shape Class}
\label{sec:hw:overview}
\begin{figure}[t]
  \centering
  \includegraphics[width=0.96\linewidth]{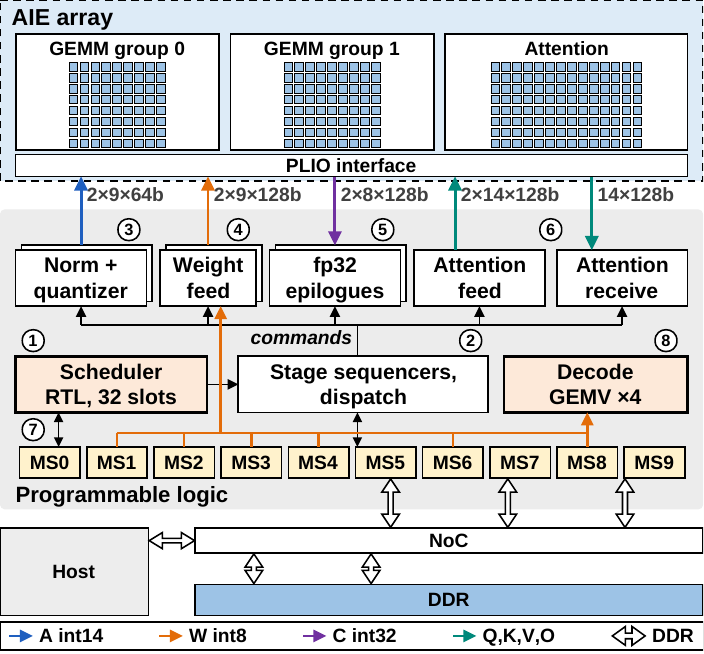}
  \caption{Architecture of \Design on the VC1902.}
  \label{fig:arch}
\end{figure}

\Design fixes its compute engines in hardware and expresses each model as programs, so one bitstream and one AIE graph serve a class of VLMs and every request can be scheduled at run time. The bitstream and the AIE graph fix a \emph{shape class}. Its LM has a hidden width (model width) of up to 2,048, head dimension 64 and up to 32 KV heads with a query-to-KV head ratio that divides 16, attention and short-convolution layers in any number and order, gated FFNs up to 10,752 wide, RMS or LayerNorm pre-norm, 1-D rotary embeddings, any vocabulary and up to 4,096 tokens of context. Its vision encoder is a SigLIP-style ViT of width up to 1,152 and head dimension 64 or 72, with $16{\times}16$ patches ($14{\times}14$ by zero padding), full attention within an image and up to six images per prompt. A program generator (Section~\ref{sec:hw:partition}) turns the descriptor of a model in this class into weights and stage programs. All models stay resident in DDR, and switching models costs a drain of the requests in flight and a kernel relaunch, without reconfiguration. In Fig.~\ref{fig:arch}, the PL units \circnum{3}--\circnum{6} feed and drain the PLIO streams of the AIE GEMM groups and the attention block, \circnum{1}, \circnum{2} and \circnum{8} schedule the work and run small decode steps, and the memory servers \circnum{7} connect these units to DDR. We use LFM2.5-VL-3B as the running example.

\subsection{Program Generator and Partition}
\label{sec:hw:partition}
\begin{figure}[t]
  \centering
  \includegraphics[width=0.96\linewidth]{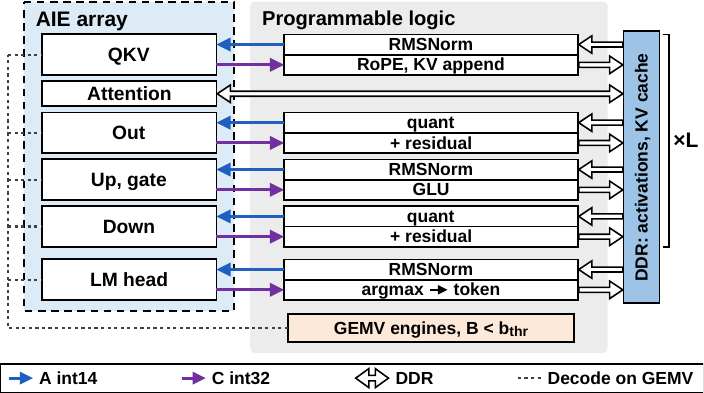}
  \caption{Partition of one attention layer of the language model and the head onto the AIE array, the PL and DDR.}
  \label{fig:partition}
\end{figure}

The program generator maps each layer around its matrix multiplies, the operators with enough reuse to keep the AIE busy, and fuses every other operator into the PL stages that feed or drain the matrix multiplies, so activations cross DDR only between matrix multiplies. It checks a model descriptor against the shape class, packs the weights into int8 with one fp32 scale per output row, fuses and places the operators, and emits checked stage programs. In Fig.~\ref{fig:partition}, each AIE operator of an attention layer and the head is bracketed by PL stages: the RMS norm and quantizer before it, and after it the epilogue with the rotary embedding and KV-cache append, the residual add, the gated activation or the argmax that yields the token. Placement follows the reuse per weight byte of Fig.~\ref{fig:vlm}(b): the encoder, projector, prefill and attention reuse each byte over many tokens and run on the AIE, whereas a decode GEMM reuses it only over the $B$ requests of a step and is marked AUTO, to be placed per step. A decode GEMM with a longer $K$ than a GEMV engine holds is pinned to the AIE. Attention, compute-bound like the GEMMs, also stays on the AIE because the PL's spare DSPs reach a small fraction of the AIE's multiply-accumulate rate and the FIFOs and epilogues already fill the PL's BRAM and URAM.

\subsection{Stage Programs and Work Units}
\label{sec:hw:prog}
A stage program describes one stage for every prompt length and batch size, so the scheduler can start, interleave and resume stages without the host. An instruction is one 64-byte DDR beat with an opcode, a placement (PL, AIE or AUTO), a token mode and typed operands. The host allocates DDR regions and writes their base addresses into a configuration block: int8 embedding rows (W8E), activation rows (XD), int8 weights in weight-feed order (W8B), programs and descriptors (CTL), per-request state and KV caches (ST), and fp32 parameters such as row scales, norms and rotary tables (P32). An address operand is a region and an offset from its base in 64-byte beats, and XD and ST offsets start at each request's own area. A scalar is a value or a register: R0 holds the rows of the stage, R1 its start position and R4 the batch $B$. Fig.~\ref{fig:prog}(a) shows the decode-step program, with hexadecimal offsets and brackets around outputs placed by a descriptor. After pcs 0--1 gather the $B$ requests' embedding rows, the epilogues of the short-convolution layer in pcs 2--5 apply the convolution (tri), the residual add (res) and the gated activation (silu, glu). In the attention layer, pc 10 applies the rotary embedding and appends to the KV cache (qkv8), and pc 11 runs attention over each request's cache in ST. Last, pc 130 selects the token (argmax) and pc 131 advances each request's position. Row $i$ of every operand belongs to request $i$, so XD+40 in pc 2 lies 0x40 beats past each request's XD base.

\begin{figure}[t]
{\scriptsize\textbf{(a)}}\vspace{-1pt}
\begin{lstlisting}[style=stageprog]
(*\pc{0}*) SETB.PL     set=R10 B=R4                ; R4 = B requests
(*\pc{1}*) ASSEMBLE.PL XD+0 <- W8E+0 seg=R9 D=2048 ; row i = request i
(*\pc{2}*) GEMM.AUTO   XD+40 <- XD+0, W8B+65f140   R4x2048x6144   rms|tri
(*\pc{3}*) GEMM.AUTO   XD+0  <- XD+40, W8B+68f140  R4x2048x2048   res
(*\pc{4}*) GEMM.AUTO   XD+80 <- XD+0, W8B+69f140   R4x2048x21504  rms|silu|glu
(*\pc{5}*) GEMM.AUTO   XD+0  <- XD+80, W8B+747140  R4x10752x2048  res
    ; ... pcs 6-9: layer 1 (conv)
(*\pc{10}*) GEMM.AUTO   [CTL+2013d] <- XD+0, W8B+8d7140  R4x2048x3072 rms|qkv8
(*\pc{11}*) MATTN.AIE   XD+40 <- XD+1d0, ST+0, ST+8000  Hq=32 Hkv=8 S0=R1
(*\pc{12}*) GEMM.AUTO   XD+0  <- XD+40, W8B+8ef140  R4x2048x2048   res
(*\pc{13}*) GEMM.AUTO   XD+80 <- XD+0, W8B+8ff140   R4x2048x21504  rms|silu|glu
(*\pc{14}*) GEMM.AUTO   XD+0  <- XD+80, W8B+9a7140  R4x10752x2048  res
    ; ... pcs 15-129: 27 more layers (8 of the 30 are attention)
(*\pc{130}*) GEMM.AUTO   [CTL+20159] <- XD+0, W8B+2aa7140 R4x2048x128000 rms|argmax
(*\pc{131}*) CTXUPD.PL   set=R10 eos=124900 maxn=R15 cpl=R13 ; pos+1, token
(*\pc{132}*) HALT
\end{lstlisting}
{\scriptsize\textbf{(b)}}\vspace{-1pt}
\begin{lstlisting}[style=aiecode]
MMUL C00, C01, C10, C11;    // 2x2 blocks, 4x4x4 int16 x int8 each
if (FIRST) { C00.mul(A0, B0); /* ... C11 */ }  // tile 0 starts the sums
else { C00 = MMUL(readincr_v<16>(ci)); /* ... */ C00.mac(A0, B0); /* ... */ }
for (unsigned i = 1; i < KB; ++i) chess_prepare_for_pipelining {
  /* load A0, A1 (int16 activations), B0, B1 (int8 weights) */
  C00.mac(A0, B0); C01.mac(A0, B1); C10.mac(A1, B0); C11.mac(A1, B1);
  nxt[0] = readincr(in); nxt[1] = readincr(in);   // next weight block
}
writeincr(co, C00.to_accum()); /* ... C11 */      // to the next tile
\end{lstlisting}
\vspace{-1ex}
\caption{(a) Decode-step program of the running example and (b) inner loop of a GEMM tile in AIE C++ (excerpts).}
\label{fig:prog}
\end{figure}

\begin{figure*}[t]
  \centering
  \includegraphics[width=0.96\textwidth]{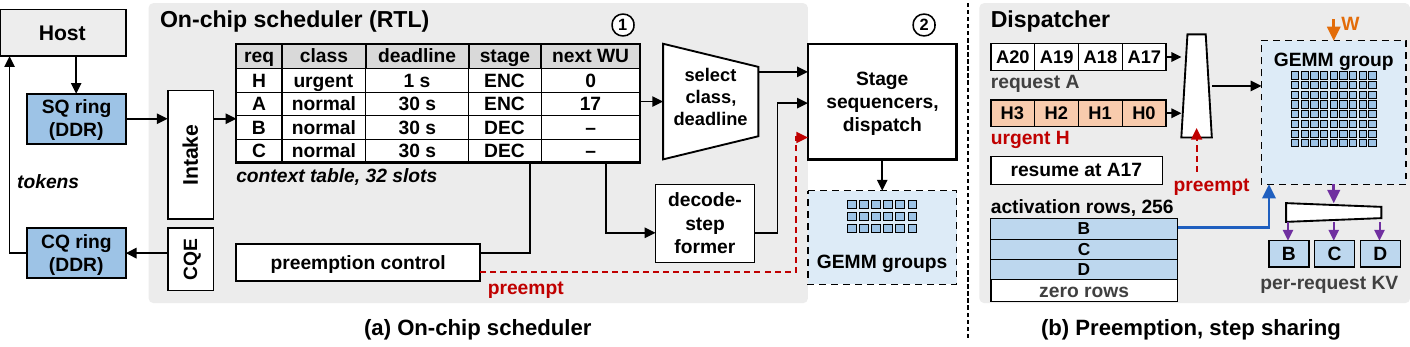}
  \caption{Multi-request execution. (a) The on-chip scheduler orders admitted requests by class and deadline. (b) An urgent request preempts at a work-unit boundary, and decoding requests share one decode step.}
  \label{fig:multireq}
\end{figure*}

Work units make the static AIE graph preemptible. A dispatcher splits each AIE instruction into \emph{work units} of at most $T_\mathrm{WU}=2$\,ms, each covering a block of rows, one $K$ segment and a range of column blocks, numbered in a fixed order that depends only on the shape. To preempt a stage, the dispatcher stops at the next boundary, drains the issued units and returns a resume record $\{$job, pc, next unit$\}$, from which the instruction is later reissued. Weights do not change and every instruction writes its results to DDR, so the resume record is the whole context of a preempted stage.

\subsection{Hardware Units}
\label{sec:hw:aie}
A GEMM group is organized so that each weight byte read from DDR serves many activation rows. Its $N_R=8$ chains have $N_T=9$ MAC tiles and one output tile each, and a tile multiplies an $m{\times}k=32{\times}128$ activation block by a $k{\times}n=128{\times}32$ weight block and passes the partial sums down the cascade, so a chain sums a $K$ segment of $N_T k=1152$ inside the array. Each weight stream is broadcast to all chains of a group, so a weight byte serves $N_R m=256$ rows, and the $G=2$ groups split the column blocks of a GEMM. A decode work unit computes only the rows that hold requests. In the inner loop of Fig.~\ref{fig:prog}(b), a tile computes its block product as $2{\times}2$ groups of $4{\times}4{\times}4$ \texttt{aie::mmul} operations. The first tile starts the sums (\texttt{mul}), and each later tile reads the 48-bit sums of its neighbor from the cascade (\texttt{ci}), adds its 128-wide slice of $K$ (\texttt{mac}) and passes the sums on (\texttt{co}). Each loaded vector feeds two multiplies, so the pipelined loop issues one 64-MAC operation per cycle while it reads the next weight block (\texttt{nxt}). Another 112 tiles, 14 columns of eight, form the attention block.

\label{sec:hw:pl}The PL units on the AIE's input and output streams take every operation that is not a dense product, so the AIE tiles stay on multiply-accumulates and the fused operations add no pass over DDR. Consider pc 2 of Fig.~\ref{fig:prog}(a). The scheduler~\circnum{1}, an RTL unit with an $N_J$-entry context table, resolves AUTO once for the decode step: the GEMV engines~\circnum{8} if $B<b_\mathrm{thr}$, the GEMM groups otherwise. If the step runs on the GEMM groups, a stage sequencer~\circnum{2} adds the region and request bases, and the dispatcher cuts work units. The norm and quantizer~\circnum{3} reads the rows at XD+0 through a memory server~\circnum{7}, applies the RMS norm and rounds the rows to int14. The weight feed~\circnum{4} streams the $2048{\times}6144$ weights from W8B+65f140 into the group's $N_T$ weight FIFOs, which are deep enough that consecutive row groups overlap in a chain. The fp32 epilogues~\circnum{5} then combine the partial sums, apply the short convolution with its P32 taps and the state in ST, and write 2048 values per request to XD+40. The attention units~\circnum{6} move queries, keys and values to the attention block and merge its 14 output streams into four. The $E=4$ GEMV engines~\circnum{8} run steps below $b_\mathrm{thr}$ and read weights through $N_P=6$ memory-server ports shared with the weight feed.

\label{sec:hw:memory}The $N_S=10$ memory servers~\circnum{7} share the DDR ports among clients with very different rates. Each server owns one port and issues a read only when the client has room for the whole burst, so a slow client does not block the others. A server keeps at least $N_\mathrm{fl}=4$\,KB in flight, but on ports that feed the AIE no more than the client buffers, so a full AIE input never stalls the shared response path.

\subsection{Integer Arithmetic}
\label{sec:method:numerics}
\Design computes every matrix multiply with integer partial sums and rounds once, in a fixed order, in the PL epilogue, so a result does not depend on the engine or on how its work units are split.

Weights are quantized after training to symmetric int8 with one fp32 scale $\sigma_c$ per output channel, equal to the channel's absolute maximum over 127. Activations are quantized per row $r$ and per scale group $g$ to int14 with scale $s_{r,g}$, equal to the group's absolute maximum over 8191, and both use round-to-nearest-even (rne):
\begin{equation}
  q_{r,k} = \mathrm{clip}\!\left(\mathrm{rne}\!\left(x_{r,k}/s_{r,g}\right), \pm 8191\right),
  \quad
  P_{r,g,c} = \sum_{k\in g} q_{r,k}\, w_{k,c}.
  \label{eq:partial}
\end{equation}
A scale group is the whole row when $K\le2048$ and one 1152-wide chain segment otherwise. Since $2048\cdot8191\cdot127 < 2^{31}$, every $P_{r,g,c}$ fits in an int32, a bound that sets the 14-bit activation width.

The fp32 epilogues then combine the scale groups and apply
\begin{equation}
  \begin{aligned}
  a_{r,c} &\leftarrow \mathrm{fmac}\!\left(P_{r,g,c},\, s_{r,g},\, a_{r,c}\right) \quad \text{for } g \text{ ascending},\\
  y_{r,c} &= \mathrm{bf16}\!\left(f(\sigma_c\, a_{r,c} + b_c)\right),
  \end{aligned}
  \label{eq:epi}
\end{equation}
where fmac is the multiply-add of the DSP58 engines in their fp32 mode (DSPFP32), which rounds the product and then the sum, and $f$ is the layer's activation, gating, convolution, rotary embedding or residual operation. The epilogue also appends to the KV cache, in int8 with one fp32 scale per layer, KV head and token (absolute maximum over 127). The AIE attention tiles compute attention in fixed 64-key blocks with integer sums, and the softmax evaluates $2^x$ with a degree-4 polynomial of single IEEE fp32 operations.

\section{Multi-Request Execution}
\label{sec:method}
\label{sec:method:multi}
Concurrent requests differ in urgency, yet their decode steps read the same weights. Building on the work units of Section~\ref{sec:hw:prog}, the on-chip scheduler of \Design exploits both. The scheduler runs on chip because a decode step issues work units for every layer, too many to launch one by one from the host. Fig.~\ref{fig:multireq}(a) follows a request: the host writes it into the submission (SQ) ring in DDR with a service class and a deadline, the intake admits it, and its tokens return to the host as completion entries (CQEs) in the completion (CQ) ring. The scheduler revisits its choice at each arrival, at the end of each work unit and at the end of each decode step.

A request is admitted when a context slot and a buffer for its intermediate data are free, and it holds one row of the context table, with its class, deadline, current stage and next work unit. In Fig.~\ref{fig:multireq}(a), H is urgent with a 1\,s deadline and waits for its encoder, A runs its encoder with next work unit 17, and B and C are decoding. The $N_J=32$ rows bound the KV-cache and state memory. A selector picks the next runnable stage by class, then by deadline, then by arrival, and the stage sequencers issue that stage's work units.

Decoding requests share one decode step, so the language-model weights are read once per step rather than once per request. Each request adds only one activation row and its own attention work, while the weight traffic stays the same. In Fig.~\ref{fig:multireq}(b), B, C and D fill three rows of one 256-row work unit, so each weight block serves all three requests, and each request writes its own KV cache. While a request still waits for its first token, every decode step runs on the PL GEMV engines, so the GEMM groups serve the prefills. Afterwards, steps of fewer than $b_\mathrm{thr}$ requests, a threshold set by the host, run on the PL and larger ones on the groups. PL steps overlap with group stages, so long stages do not stall decoding.

Preemption acts at work-unit boundaries and costs only a resume record. In Fig.~\ref{fig:multireq}(b), the urgent request H becomes runnable while the GEMM groups run the encoder of request A, and a projector or prefill is handled alike. The dispatcher finishes the running work unit, at most 2\,ms long, lets the up to two attention calls already issued complete, and then issues H0 to H3, so preemption takes effect within one work unit and the attention calls in flight. Request A keeps its results in DDR, and its table row records A17, where A resumes.

\section{Evaluation}
\label{sec:experiments}

\subsection{Experimental Setup}
\label{sec:exp:setup}
We evaluate \Design on an AMD Versal VC1902 (Vitis 2022.1, TAPA~\cite{tapa}) at 240\,MHz in the PL and 1.25\,GHz in the AIE. We serve LFM2.5-VL-3B with the integer arithmetic of Section~\ref{sec:method:numerics} on four kinds of requests: \reqnum{1} a 177-token camera question (one 512$\times$288 image: 144 image and 33 text tokens), \reqnum{2} a 60-token text question, \reqnum{3} a 555-token three-image question (three 512$\times$288 images: 432 image and 123 text tokens) and \reqnum{4} a 4,088-token text question.

As GPU baselines, we use an NVIDIA Tesla T4 of similar power to the VC1902, the edge Jetson Orin Nano Super and the high-end desktop RTX 3090 Ti (Table~\ref{tab:platform}(a)), with 2 to 20 times the VC1902's memory bandwidth. They run the same model, prompts and int8 weights with CUDA graphs and optimized Triton kernels from torch.compile. We also compare with an all-AIE build that runs every operation on the AIE, including the normalizations, quantization, softmax and fp32 epilogues that \Design runs in the PL; there, the PL only moves data.
\newsavebox{\tabplat}\newsavebox{\tabacc}
\begin{table}[t]
\centering
\caption{Platform specs and accuracy.}
\label{tab:platform}\label{tab:accuracy}
\scriptsize
\renewcommand{\arraystretch}{0.72}\setlength{\aboverulesep}{0.1ex}\setlength{\belowrulesep}{0.25ex}\setlength{\tabcolsep}{1.6pt}
\sbox{\tabplat}{\begin{tabular}[t]{@{}lcr@{}}
\toprule
\textbf{Platform} & \textbf{Memory} & \textbf{GB/s} \\
\midrule
VC1902 & DDR4 & 50 \\
T4 & GDDR6 & 320 \\
Orin Nano Super & LPDDR5 & 102 \\
RTX 3090 Ti & GDDR6X & 1,008 \\
\bottomrule
\end{tabular}}\sbox{\tabacc}{\begin{tabular}[t]{@{}lrcc@{}}
\toprule
\textbf{Benchmark} & \textbf{N} & \textbf{fp32} & \textbf{Int.} \\
\midrule
RealWorldQA~\cite{realworldqa} & 765 & 69.15\% & 69.67\% \\
LingoQA~\cite{lingoqa} & 100 & 54.0\% & 54.0\% \\
LongBench-E~\cite{longbench} 3--4k & 149 & 61.2 & 60.8 \\
PG-19~\cite{pg19} ppl.\ 4k & 50 & 19.17 & 19.20 \\
\bottomrule
\end{tabular}}\subfloat[Platform specs.]{\begin{minipage}[c][\dimexpr\ht\tabacc+\dp\tabacc][t]{\wd\tabplat}\usebox{\tabplat}\end{minipage}}\hfill
\subfloat[Accuracy.]{\begin{minipage}[c][\dimexpr\ht\tabacc+\dp\tabacc][t]{\wd\tabacc}\usebox{\tabacc}\end{minipage}}
\end{table}
\begin{figure}[t]\centering\includegraphics[width=0.9\columnwidth]{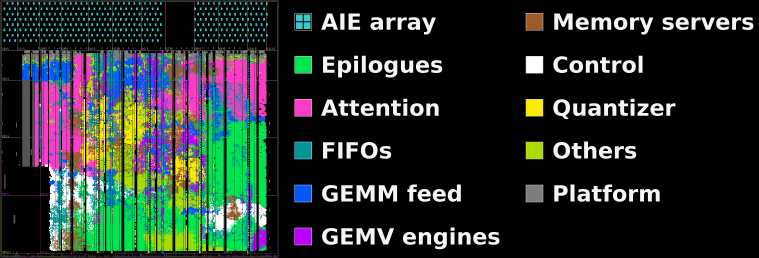}\vspace{-1ex}\caption{Placed and routed design  on the VC1902, colored by unit.}\label{fig:floorplan}\end{figure}

\begin{table}[t]
\centering
\caption{Resources of the routed build by unit class.}
\label{tab:resources}
\renewcommand{\arraystretch}{0.72}\setlength{\aboverulesep}{0.1ex}\setlength{\belowrulesep}{0.25ex}
\scriptsize
\setlength{\tabcolsep}{0pt}
\begin{tabular*}{\columnwidth}{@{\extracolsep{\fill}}lrrrrr@{}}
\toprule
\textbf{Unit class} & \textbf{LUT} & \textbf{FF} & \textbf{BRAM36} & \textbf{URAM} & \textbf{DSP} \\
\midrule
Epilogues & 134.9k & 179.9k & 109 & 160 & 430 \\
Attention & 102.3k & 97.0k & 24 & 24 & 35 \\
FIFOs & 61.2k & 111.7k & 473 & 148 & 0 \\
GEMM feed & 40.3k & 38.1k & 1 & 0 & 16 \\
GEMV engines & 36.3k & 44.5k & 109 & 32 & 456 \\
Memory servers & 33.1k & 33.4k & 0 & 0 & 0 \\
Control & 40.9k & 43.8k & 24 & 1 & 38 \\
Quantizer & 19.3k & 15.2k & 16 & 48 & 273 \\
Others & 51.3k & 50.1k & 73 & 8 & 158 \\
Platform & 31.1k & 49.7k & 32 & 0 & 0 \\
\midrule
Total & 550.7k & 663.4k & 861 & 421 & 1,406 \\
\% of the VC1902 & 61.2\% & 36.9\% & 89.0\% & 90.9\% & 71.4\% \\
\bottomrule
\end{tabular*}
\end{table}

\begin{table}[t]
\centering
\caption{Decode of one request on \Design, with all on the AIE, and on the GPUs.}
\label{tab:decode}
\renewcommand{\arraystretch}{0.72}\setlength{\aboverulesep}{0.1ex}\setlength{\belowrulesep}{0.25ex}
\scriptsize
\setlength{\tabcolsep}{3.2pt}
\begin{tabular}{@{}lcccc@{}}
\toprule
\textbf{Decode on} & \textbf{ms/token} & \textbf{Tokens/s} & \textbf{J/token} & \textbf{W} \\
\midrule
All on the AIE (no PL compute) & 154.4 & 6.5 & 8.45 & 54.7 \\
Tesla T4 (int8 weights) & 87.0 & 11.5 & 5.83 & 67.0 \\
Jetson Orin Nano Super (int8 weights) & 268.2 & 3.7 & 4.91 & 18.3 \\
RTX 3090 Ti (int8 weights) & 18.4 & 54.3 & 6.96 & 378 \\
\midrule
\Design (PL GEMV engines) & 66.5 & 15.0 & 3.56 & 53.6 \\
\bottomrule
\end{tabular}
\end{table}

\subsection{Accuracy}
\label{sec:exp:acc}
On the full RealWorldQA set (Table~\ref{tab:accuracy}(b)), the integer arithmetic scores 0.52 points above fp32, with a paired 95\% confidence interval of $\pm0.63$ points. At the 4,096-token maximum context, it predicts the same next token as fp32 at 98.0\% of positions in 50 \mbox{PG-19} books.

\subsection{Resource Utilization}
\label{sec:exp:hw}
In Table~\ref{tab:resources}, the epilogues and attention units of Fig.~\ref{fig:floorplan} use the most LUTs, the FIFOs the most BRAM, the GEMV engines and epilogues the most DSPs, and the epilogues and FIFOs the most URAM. The AIE array uses 272 of its 400 tiles: 160 in the two GEMM groups (nine MAC and one output tile per chain) and 112 for attention.

\subsection{Latency and Throughput}
\label{sec:exp:e2e}
When idle, a text question~\reqnum{2} gets its first token in 140\,ms, 133\,ms in the prefill. A camera question~\reqnum{1} gets it in 303\,ms (130\,ms encoder, 165\,ms prefill), and a three-image question~\reqnum{3} in 838\,ms. Camera and text first tokens come 1.7 and 2.0 times sooner than on the Orin Nano Super. With the GEMM groups and the attention block working together, the 4,088-token prefill~\reqnum{4} delivers 6.0 useful TOPS.

\begin{figure}[t]
  \centering
  \includegraphics[width=\linewidth]{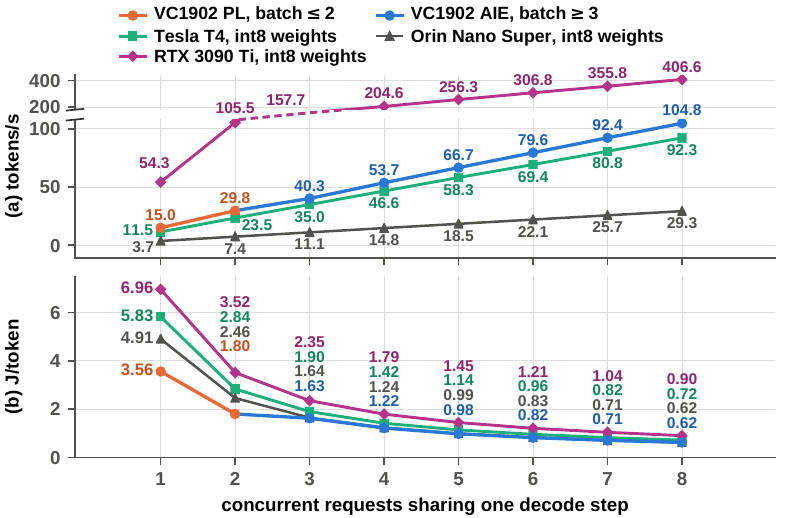}
  \caption{Decode with shared steps at 240\,MHz once no request waits for its first token (PL for batch ${\leq}2$, two GEMM groups for ${\geq}3$), and on the T4, RTX 3090 Ti and Orin Nano Super: (a) throughput, (b) energy per token.}
  \label{fig:tput}
\end{figure}

On the PL (Fig.~\ref{fig:tput}), two requests share one 67.2\,ms step, so throughput doubles to 29.8 tokens/s. Each weight beat serves two requests per cycle, so three take about as long as four. On the two GEMM groups, with epilogues on the PL, a step stays between 73.3 and 76.3\,ms from one to eight requests, while a PL step grows from 66.5 to 185.5\,ms. The two paths cross between two and three requests, and beyond that the groups take up to 59\% less time. With the split of Section~\ref{sec:method}, which runs every step on the PL until each request has its first token and uses $b_\mathrm{thr}=3$ afterwards, eight text requests of 64 output tokens arriving together finish in 6.25\,s, against 6.34 and 6.96\,s with fixed thresholds of seven and three, 12.44\,s all on the PL and 12.85\,s on the all-AIE build. A shared step of eight requests delivers 104.8 tokens/s, against 92.3 on the T4 and 29.3 on the Orin Nano Super. Energy per token is below the T4's at every batch size: 1.22 against 1.42\,J at four requests, 0.62 against 0.72\,J at eight.

\subsection{Multi-Request Execution}
\label{sec:exp:multi}
In Table~\ref{tab:preempt}, an urgent camera question~\reqnum{1} arrives while a normal request runs its encoder or prefill, e.g., 49.9\,ms into a normal camera question. Entries are relative to each platform's idle first token: 303, 156, 518 and 41\,ms on \Design, the T4, Orin Nano Super and RTX 3090 Ti. With preemption, the urgent first token stays at 1.00 to 1.05 times the idle case. When the urgent question arrives midway through a 4k-token prefill~\reqnum{4}, its first token comes 14\,ms later than when idle. With priority only, the urgent question waits for the running stage, and in arrival order it also waits for the normal request's first token. Behind a 4k-token question, both policies miss a 1\,s deadline, and arrival order also misses it behind three images. The normal request resumes at its saved work unit, with its first token later than in arrival order by the urgent question's 303\,ms plus 17 to 153\,ms. Decode continues during preemption: in a run of three text requests where one prefill pauses for another, 8 of its 11 decode steps run on the PL while the groups serve the prefills. The GPUs preempt between stages and between 512-token prefill chunks, so their urgent first token comes 1.10 to 2.39 times their idle case, against 1.00 to 1.05 on \Design.

Preemption also keeps urgent deadlines under load. Camera~\reqnum{1}, text~\reqnum{2} and three-image~\reqnum{3} questions arrive as a Poisson process at 0.05 to 0.45 requests/s, each load point pooling ten 600\,s traces. A quarter are urgent camera questions with a 1\,s first-token deadline, the rest normal with 32 to 128 output tokens. One context slot is kept for urgent requests. At 0.3 requests/s (410 urgent, 1,292 normal), no urgent request is late (p99 first token 0.61\,s). With priority only, none is late (p99 0.97\,s), and in arrival order 4.2\% (1.34\,s). If one normal request in ten is a 4k-token question~\reqnum{4}, the urgent p99 is 0.62\,s at 0.2 requests/s, against 4.41\,s with priority only and 4.76\,s in arrival order.

\begin{table}[t]
\centering
\caption{First-token time of an urgent camera question when another request runs ahead of it, relative to its idle first-token time.}
\label{tab:preempt}
\renewcommand{\arraystretch}{0.72}\setlength{\aboverulesep}{0.1ex}\setlength{\belowrulesep}{0.25ex}
\scriptsize
\setlength{\tabcolsep}{0pt}
\begin{tabular*}{\columnwidth}{@{\extracolsep{\fill}}lccc@{}}
\toprule
 & \multicolumn{3}{c}{\textbf{Request ahead}} \\
\cmidrule(l){2-4}
\textbf{Scheduling} & \textbf{Camera}~\reqnum{1} & \textbf{3 images}~\reqnum{3} & \textbf{4k text}~\reqnum{4} \\
\midrule
\textbf{\Design: preemption} & \textbf{1.00$\times$} & \textbf{1.00$\times$} & \textbf{1.05$\times$} \\
\Design: priority only & 1.26$\times$ & 2.18$\times$ & 6.63$\times$ \\
\Design: arrival order & 2.05$\times$ & 3.75$\times$ & 6.75$\times$ \\
\cmidrule{1-4}
T4: preemption & 1.11$\times$ & 1.65$\times$ & 1.33$\times$ \\
T4: arrival order & 1.93$\times$ & 3.22$\times$ & 8.48$\times$ \\
\cmidrule{1-4}
Orin Nano Super: preemption & 1.10$\times$ & 1.48$\times$ & 2.26$\times$ \\
Orin Nano Super: arrival order & 1.88$\times$ & 3.04$\times$ & 6.76$\times$ \\
\cmidrule{1-4}
RTX 3090 Ti: preemption & 1.15$\times$ & 1.77$\times$ & 2.39$\times$ \\
RTX 3090 Ti: arrival order & 1.87$\times$ & 3.30$\times$ & 8.34$\times$ \\
\bottomrule
\end{tabular*}
\end{table}

\begin{table}[t]
\centering
\caption{Ablation at 240\,MHz. Times in ms; group decode step (PL epilogues kept) with one request, tokens/s with three.}
\label{tab:ablation}
\renewcommand{\arraystretch}{0.72}\setlength{\aboverulesep}{0.1ex}\setlength{\belowrulesep}{0.25ex}
\scriptsize
\setlength{\tabcolsep}{2pt}
\begin{tabular*}{\columnwidth}{@{\extracolsep{\fill}}lccc@{}}
\toprule
\textbf{Configuration} & \textbf{First token} & \textbf{Step} & \textbf{Tok/s} \\
\midrule
1 GEMM group, AIE decode & 882 & 340.3 & 8.8 \\
2 GEMM groups, AIE decode & 724 & 208.7 & 14.3 \\
Deep attention buffers & 469 & 208.7 & 14.3 \\
Row-scaled AIE decode & 447 & 121.2 & 24.6 \\
Attention-call overlap & 415 & 121.2 & 24.6 \\
PL decode & 415 & 121.2 & 29.7 \\
Attention output + GEMM feed & 356 & 90.5 & 32.8 \\
\textbf{Balanced GEMM split + attention write} & \textbf{303} & \textbf{73.3} & \textbf{40.3} \\
\bottomrule
\end{tabular*}
\end{table}

\subsection{Ablation Study}
\label{sec:exp:ablation}
Table~\ref{tab:ablation} adds the design choices row by row. A second GEMM group splits each GEMM's column blocks, cutting the first token to 724\,ms and a group decode step to 208.7\,ms.
Deep attention buffers hold a whole pass, so all 14 columns work at once instead of about one, cutting the encoder from 487 to 232\,ms.
Row-scaled AIE decode, computing only the rows in use and prefetching the next weights, cuts a decode step to 121.2\,ms and, with a shorter prefill and output layer, the first token to 447\,ms. Overlapping attention calls cuts the encoder from 232 to 203\,ms and the first token to 415\,ms. PL decode for small steps raises the throughput of three requests from 24.6 to 29.7\,tokens/s. Pipelining the attention output and merging its 14 streams into four cuts the encoder from 203 to 157\,ms. Sending each work unit's zero rows before its activation rows in the GEMM feed cuts a group step to 90.5\,ms and the first token to 356\,ms. Last, splitting each GEMM's column blocks evenly between the two groups, issuing a long-$K$ GEMM one $K$ segment at a time so the quantizer prepares each segment once, and pipelining the attention write cut the encoder from 157 to 130\,ms, the camera prefill from 192 to 165\,ms, a group step from 90.5 to 73.3\,ms and the first token from 356 to 303\,ms.

\subsection{Single-Request Decode}
\label{sec:exp:compare}
Table~\ref{tab:decode} compares one-request decode, the memory-bound phase that the other accelerators of Table~\ref{tab:related} do not run. The PL takes 66.5\,ms per token, 2.3, 1.3 and 4.0 times faster than all on the AIE, the T4 and the Orin Nano Super, reading the weights at 40.6\,GB/s, 81\% of the DDR ceiling of about 50\,GB/s. Each token uses 61\%, 73\% and 51\% of the energy of the T4, Orin Nano Super and RTX 3090 Ti.

\subsection{Multi-Model Support}
\label{sec:exp:model2}
The same build serves two other VLMs of its shape class (Table~\ref{tab:model2}): LFM2.5-VL-1.6B, with the 3B's vision encoder and 16 instead of 30 LM layers, and SmolVLM2-2.2B, from another family, with 14-pixel patches and 32 KV heads. The two models score 61.6\% and 58.3\% on RealWorldQA, against 62.0\% and 57.9\% in fp32. A switch between the three resident models takes a median of 33\,ms.
\begin{table}[!t]
\centering
\caption{Two other VLMs on the same build.}
\label{tab:model2}
\renewcommand{\arraystretch}{0.72}\setlength{\aboverulesep}{0.1ex}\setlength{\belowrulesep}{0.25ex}
\scriptsize
\setlength{\tabcolsep}{3pt}
\begin{tabular}{@{}lcc@{}}
\toprule
 & \textbf{LFM2.5-VL-1.6B} & \textbf{SmolVLM2-2.2B} \\
\midrule
First token, text / camera & 63 / 209\,ms & 106 / 269\,ms \\
PL decode, one request & 29.5\,ms & 46.6\,ms \\
Tokens/s, one / eight requests & 33.9 / 215.3 & 21.4 / 101.1 \\
Energy, one / eight requests & 1.57 / 0.30\,J/token & 2.65 / 0.57\,J/token \\
Weight read rate, of DDR ceiling & 39.6\,GB/s, 79\% & 36.7\,GB/s, 73\% \\
Resident weights & 1.78\,GB & 2.29\,GB \\
\bottomrule
\end{tabular}
\end{table}

\section{Conclusion}
\label{sec:conclusion}

A static AIE dataflow graph can serve dynamic, deadline-driven VLM traffic when the programmable logic owns control and memory. In \Design, the reuse of each weight byte decides where a phase runs: phases that reuse a byte across many tokens fill the AIE array, while decode steps of a few requests reuse each byte too little and run in the PL next to the DDR ports. Short work units make preemption cheap. A preemption waits for at most one work unit and the attention calls in flight, and it saves no on-chip state because every unit leaves its results in DDR, so an urgent camera question gets its first token within 1.05 times its idle time. Binding the model at run time lets one bitstream serve two other VLMs of the same shape class with their own programs and weights.

Each bitstream serves one shape class, and the same flow builds a new bitstream from the descriptor of any model outside that class, such as a larger model or one with another layer type. Decode follows the DDR bandwidth of the VC1902, so Versal AIE SoCs with HBM and AIE-ML tiles would raise the decode rate and attention throughput.

\bibliographystyle{IEEEtran}
\bibliography{references}

\end{document}